# A Study on Anxiety about Using Robo-taxis:

# HMI Design for Anxiety Factor Analysis and Anxiety Relief Based on Field Tests

Soyoung Yoo[1], Sunghee Lee[1], Seongsin Kim[1], Eunji Kim[1], Hwan Hwangbo[2], Namwoo Kang[1,*]

[1]Mechanical Systems Engineering, Sookmyung Women's University

[2]Hyundai Motor Company

*Corresponding author: nwkang@sm.ac.kr

## Abstract

Despite the approaching commercialization of robo-taxis, various anxiety factors concerning the safety of autonomous vehicles are expected to form a large barrier against consumers' use of robo-taxi services. The purpose of this study is to derive the various internal and external factors that contribute to the anxieties of robo-taxi passengers, and to propose a human-machine interface (HMI) concept to resolve such factors, by testing robo-taxi services on real, complex urban roads. In addition, a remote system for safely testing a robo-taxi in complex downtown areas was constructed, by adopting the Wizard of Oz (WOZ) methodology. From the results of our tests – conducted upon 28 subjects in the central area of Seoul – 19 major anxiety factors arising from autonomous driving were identified, and seven HMI functions to resolve such factors were designed. The functions were evaluated and their anxiety reduction effects verified. In addition, the various design insights required to increase the reliability of robo-taxis were provided through quantitative and qualitative analysis of the user experience surveys and interviews.

**Keywords:** Robo-taxi, Self-driving car, Anxiety, Human-machine interface, Field test

# 1. Introduction

Waymo, a subsidiary of Alphabet, recently launched the world's first commercial robo-taxi service (LA Times, 2018), and formed a partnership with Renault-Nissan-Mitsubishi. Together they are developing the robo-taxi services (reservation and payment systems) required to produce and operate robo-taxis equipped with the autonomous driving technology of Waymo, by 2022 (Nikkei Asian Review, 2019). Daimler-Benz have signed a contract with Bosch, a company specializing in components, and will start road-testing complete robo-taxis in San Jose, USA. It will also provide car-sharing services by developing dispatch applications (Forbes, 2018). Tesla is preparing a robo-taxi service in which vehicles not being used by their owners can be instead used as robo-taxis, with the profits being shared between the owners and Tesla (USA Today, 2019). As described, major global companies are accelerating their service innovations, by combining car-sharing and autonomous driving technologies.

In South Korea, the world's first 5G transmission began at the end of 2018, ushering in a communication era 20 times faster than before (Rcrwirelessnews, 2019). In addition, connected-car and car-sharing services are being actively developed alongside these technological advances. In particular, Sejong City designated "free zone" to regulate self-driving cars, and introduced a test bed of self-driving buses for citizens residing in offroad areas. Sejong City aims to gradually increase the number of service areas and commercialize them in 2022 (Kang, 2019). Moreover, SKTelecom constructed a test bed in Seoul City, and set up infrastructures such as control platforms and 5G networks in Sangam Digital Media City (DMC), so as to conduct tests on autonomous driving technologies (Koreatimes, 2019). The government also designated regulation-free zones, including Hwaseong K city, where autonomous vehicles can be tested on ordinary roads, so as to promote active research and development (Koreajoongangdaily, 2019). Nevertheless, experimental data in South Korea are noticeably deficient compared to that of the USA, where tests on autonomous vehicles are permitted on real public roads (Engadget, 2019). This is because most previous tests of autonomous vehicles in South Korea were conducted in controlled environments or in specially fabricated test beds (Forbes, 2017).

The commercialization of autonomous vehicles is approaching, and related services will bring many benefits (Kang et al., 2017; Lee et al., 2019), however, issues surrounding their reliability and safety are becoming major barriers against consumers' adoption of autonomous vehicles (Bansal et al., 2016; Schoettle, & Sivak, 2014). Reliability is known to be one of the most important factors affecting users' acceptance of automated systems (Carter & Bélanger, 2005). Although research on reliability is well-established in the field of psychology, studies on the reliability of autonomous vehicles, as perceived by humans, are insufficient (Stanton et al., 2000; Shahrdar, Menezes, & Nojoumian, 2018).

Shahrdar et al. (2019) assessed passenger perceptions of the reliability of autonomous vehicles in real time, using a virtual reality (VR) simulator, they then analyzed the factors that increase or decrease this perceived reliability. Lee et al. (2016) tested a prototype autonomous vehicle after constructing a test bed on a road within a university campus. They showed that anxieties over unpredictable situations can be reduced to some extent if passengers are made aware of the fact that the autonomous vehicle is in proper operation and can make accurate decisions in various driving situations. However, these tests use only a simulator or test bed, and thus have only a limited capacity to investigate the various factors affecting the perceived reliability of autonomous vehicles. If the negative empirical data of users in real road situations can be obtained in advance, problems surrounding the commercialization of robo-taxis can be predicted and reflected in designs. Therefore, there is a growing demand for studies based on field tests (Ghazizadeh et al., 2012).

To overcome the challenges of autonomous vehicle field tests, studies have been conducted using the Wizard of Oz (WOZ) methodology. WOZ is a method in which a tester plays the role of the automated system from behind a curtain, even though a service that can actually operate the system is provided (Dahlbäck et al., 1993; Maulsby et al., 1993). Kim et al. (2020) implemented a robo-taxi service on real roads using WOZ and proposed a service design method to resolve pain points and to strengthen the elements that positively affected satisfaction with the service. Rothenbucher et al. (2016) conducted research upon the interactions between unmanned vehicles, pedestrians, and drivers on ordinary roads using WOZ.

The purpose of this study is to determine the major factors that contribute to user anxieties, by testing robo-taxi services on real, complex public road networks, and to propose a human-machine interface (HMI) design that can improve reliability using these factors. The WOZ method was adopted here in a way identical to that of Kim et al. (2020), so as to test anxiety using actual service situations in downtown areas, where taxis are most frequently used. The subjects participated in the tests whilst under the belief that the services were actual robo-taxi services, and the interactions between the subjects and robo-taxis were remotely controlled from a control tower.

Figure 1 shows the research framework of this study, which consists of four steps: test design, $1^{st}$ field test, HMI design, and $2^{nd}$ field test. In the test design step, scenarios known to cause anxiety were created, and driving paths were designed accordingly. In addition, a robo-taxi, service application, and control tower system were constructed for the field tests. In the $1^{st}$ field test, the test results were analyzed and key anxiety factors were identified. The test was divided into the pilot and main tests. The test environment was examined by vehicle development experts and non-experts during the pilot test, and shortcomings were modified and reflected in the main test. In the HMI design step, HMI solutions to resolve the key factors were found and designed. Finally, in the $2^{nd}$ field test we examined whether the newly developed HMI was able to reduce the anxieties induced by the robo-taxi. Similar to the $1^{st}$ field test, the main test was conducted after the pilot.

In the remaining sections of this paper, each of the four steps of the research framework are introduced in detail. Section 2 introduces the method of designing the robo-taxi anxiety test, and Section 3 analyzes the results and anxiety factors obtained through the $1^{st}$ field test. Section 4 proposes HMI solutions to resolve such factors, and Section 5 analyzes the results of the $2^{nd}$ field test as well as the effects of the HMI solutions. Finally, Section 6 summarizes the insights obtained with the experimental results and suggests future studies.

[Figure 1 near here]

## 2. Test Design

In this step, a number of scenarios known to contribute to autonomous vehicle-induced anxieties were created, and a survey was conducted upon the degree of anxiety expected for each scenario. Based on these results, real driving paths were selected to reflect the major scenarios. In addition, a robo-taxi to be used in field tests, an in-vehicle display and mobile application for passenger interactions, and a control tower system to perform WOZ were implemented. Finally, the interviews to be conducted before and after each field test, as well as the relevant questionnaires, were designed.

### *2.1. Design of anxiety scenarios and routes*

There are 84 scenarios that may cause anxiety, arising from the driving environment, road conditions, driver type, and vehicle controls, these problem were selected through a brainstorming session. In addition, the scenarios were classified depending on whether they could be resolved by HMI designs or only by sensors and improved autonomous driving technologies. The scenarios that belonged to the latter group were excluded, because the purpose of this study was to design a HMI that reduces anxiety. Similar factors were then grouped together, resulting in 33 factors, then 22 scenarios that could be implemented through field tests were selected.

For the 22 scenarios selected, an online survey evaluating – on a 7-point Likert scale – the degree of anxiety expected for each scenario was conducted, with 320 ordinary (non-expert) respondents. In this case, as the online respondents had no experience of robo-taxis, they were told to evaluate their anxiety assuming that, instead of being in a robo-taxi, they were in the passenger seat of a vehicle driven by a friend who had recently acquired a driver's license, for an effective survey. Of the respondents 70.7% were male and 29.3% were female, 12.9% of them were in their twenties or younger, 16.2% in their thirties, 49.2% in their forties, 18.1% in their fifties, and 3.6% in their sixties. Based on the evaluation results, Table 1 shows the top 11 scenarios out of 22 scenarios in which an average anxiety of 5 points or higher was expected were finally selected, and real roads that could reflect them were surveyed.

Various routes in which the scenarios were likely to occur were derived using the road view of an online map, and the final test path was determined by visiting them. The final driving path started from Sookmyung Women's University and returned to the university via Gyeonggyojang, it required approximately 60 minutes of driving. Figure 2 shows the driving path, and Table 1 summarizes the anxiety scenarios that are likely to occur in each section. In actual tests, scenarios other than the selected top 11 scenarios could also occur on the test path.

[Figure 2 near here]

[Table 1 near here]

Section S1 contained a two-lane road and a right-turn road in front of a school. It was chaotic due to the frequent jaywalking of pedestrians and the presence of large numbers of motorcycles. In addition, it was frequently observed that many porters unloaded objects from trucks directly onto the road. Therefore, it was expected that the scenarios of "when a pedestrian suddenly jumps out from a crosswalk in the driving signal (1$^{st}$ ranking)", "when a large object falls from a truck in the middle of the road (2$^{nd}$ ranking)", and "when a motorcycle is running next to the vehicle (9$^{th}$ ranking)" could be implemented in this section.

Section S2 passes in front of a train station. It was crowded due to a high concentration of taxis and passengers, as well as the many trucks and buses that cross the intersection. Therefore, it was expected that "when the distance to the vehicle ahead is very small (5$^{th}$ ranking)", "when the view is blocked by a large vehicle ahead (8$^{th}$ ranking)", and "when trying to change lanes in a congested area (11$^{th}$ ranking)" could be implemented. Section S3 passes intersections consecutively and includes turns. In this section, the occurrence of "when turning at high speed without slowing down (3$^{rd}$ ranking)" was expected. Moreover, an 'accident notification application' was developed and artificial accident situations were created to observe "when turning left with the right blinker on (malfunction) (6$^{th}$ ranking)", "when there is a strange noise in the vehicle (7$^{th}$ ranking)", and "when stopping in the middle of a crosswalk (10$^{th}$ ranking)" at once. To simulate such situations, Section S4, which runs in front of an apartment complex where vehicles can travel at low speed, was selected. Finally, Section S5, an alleyway, was added to the path to examine "when traveling at high speed in a narrow alley (4$^{th}$ ranking)".

## 2.2. Experimental environment design

### 2.2.1. Smartphone Application and Interaction Display

The robo-taxi service comprised four steps: (1) calling, (2) boarding, (3) traveling, and (4) disembarking. For the services required at each step, a smartphone app (an app for calling) and an in-vehicle interaction display were developed. Figure 3 shows representative screens of the developed software. In the calling step, the user enters a starting point and a destination using the smartphone app, this requests the taxi. In the boarding step, the user checks the pick-up location of the taxi using the smartphone app, and boards the taxi using the vehicle information and key, both provided through the app. When boarding is completed, basic comments ("hello, welcome to the robo-taxi" and "please fasten your seat belt for a safe trip") are heard from the display installed in front of the seat. When the user fastens the seat belt, the vehicle starts traveling and operates the navigation system to show the travel path. Finally, in the disembarkation step, when the taxi arrives at the destination, the operation of the navigation system stops and the user gets out of the vehicle. Only the most basic functions were applied to the interaction display in the 1$^{st}$ field test, and specific HMI functions were applied in the 2$^{nd}$ field test.

[Figure 3 near here]

Furthermore, in the 1$^{st}$ field test, an accident alarm function was applied to the interaction display, and the user response was tested to implement "when there is a strange noise in the vehicle (7$^{th}$ ranking)" from the anxiety scenarios in Table 1. This function notifies passengers when an accident has occurred in the path ahead of the traveling autonomous vehicle. A standby button for the vehicle system – to allow the passenger to halt the vehicle, and a control tower connection button – to activate a phone call to a service representative, were created. Figure 4 shows how to use the accident occurrence alarm function.

[Figure 4 near here]

### 2.2.2. Robo-taxi implementation

The robo-taxi interior and exterior environments were constructed based on the passenger car model most popularly employed for ordinary taxis in South Korea (see Figure 5). In the driver's seat, an opaque bulkhead was installed to completely separate the passenger in the passenger seat from the driver's seat. The passenger was told that an autonomous driving technician was on board for an emergency. In accordance with the WOZ methodology, the driver in the driver's seat actually drove the vehicle during the tests. To resolve the view problem caused by the bulkhead, cameras were installed on the side and rear of the vehicle so that the driver could see the outside through the display. In addition, communication with the control tower was enabled.

[Figure 5 near here]

To monitor and record the field test situations in real time, the images of the user, external road situations, and vehicle speed data were collected. To this end, a camera was installed in the passenger seat. In addition, a 360-degree camera was installed outside the vehicle, while On-Board Diagnostics II (OBD2) was installed inside the vehicle. Moreover, a Wi-Fi terminal was installed inside the vehicle, to enable remote control of all functions from the control tower, all cameras were connected to a smartphone, and this smartphone was connected to the internet so that it could be remotely accessed from the control tower. With this set-up, the status of the passenger and the driving situation could be observed in real time, and the information required by the user could be provided through the display. The interaction display was installed in front of the passenger seat.

One interaction display was used in the 1$^{st}$ field test, whereas four interaction displays were used in the 2$^{nd}$ field test due to the addition of new HMI functions. The details will be introduced in Section 4.

*2.2.3. Control Center*

The researcher operating the control tower provided the driver with voice instructions on the vehicle control, the driver listened to the instructions through an earphone and carried them out. The display in the passenger seat was directly controlled by the control tower, and it interacted with the user. Figure 6 shows the control tower environment and the monitoring for the driving situation in the 1$^{st}$ field test.

[Figure 6 near here]

In the 2nd field test, the control tower and the driver were required to perform more controls, to implement the new HMI functions. In particular, the control tower typed out response comments in real time, to be implemented by an artificial intelligence speaker, and the comments were converted into voice and delivered to the in-vehicle speaker (see Figure 7(a)). Moreover, for assurances at dangerous moments during driving, the driver could transmit comments using an app developed in advance (see Figure 7(b)). For example, when a certain button was clicked, the message "a congested section is recognized. I will drive safely." was transmitted. A detailed description of the functions will be provided in Section 4.

[Figure 7 near here]

## 2.3. Design of image-based interviews and surveys

Interviews and surveys were conducted before and after each field test. In the surveys and interviews conducted before each field test, the subject's knowledge of and trust in autonomous driving were examined. In the surveys and interviews conducted after each field test, the focus was that of deriving the anxiety factors felt by subjects while they used the robo-taxi. Details of the questions can be found in the appendix. In particular, the interviews given after the field tests were conducted while watching recorded videos of the subjects in the vehicle. In this study, a smartphone clicker app was created to record moments when the subjects felt anxious. The clicker app in the smartphone used to call the vehicle was automatically executed from the moment the vehicle began to move to the moment when the subject got off, and the subjects were told to click whenever they felt anxious. During the video interviews, the vehicle interior/exterior videos of the moments when the participants pressed the clicker were shown, and the reasons for pressing the clicker and the degree of anxiety were queried. Figure 8 shows a scene of the interview using the clicker.

[Figure 8 near here]

The number of participants used in the field test was 28 in the 1st pilot test, 18 in the 1st main test, 1

in the 2nd pilot test, and 6 in the 2nd main test. Of the participants, 25% were male and 75% were female, 50% of them were in their twenties, 18% in their thirties, 18% in their forties, and 14% in their fifties or older. As for the driving experience, 21% of the participants had no experience, 39% had less than 3 years' experience, 4% had 3-6 years, 7% had 6-10 years, and 29% had more than 10 years' experience; 61% of them drove less than once a week, 7% drove once a week, 18% 2-3 times a week, 7% 4-6 times a week, and 7% every day. For the frequency of taxi usage, 54% of them took a taxi less than once a week, 36% once a week, and 11% 2-3 times a week.

The interviews conducted before the field tests showed that the anxiety felt in vehicles driven by other people was, on average, 3.04 points on a 7-point scale. The average anxiety felt in ordinary taxis was 3.64 points on a 7-point scale, indicating that people felt higher anxiety compared to when they were in ordinary vehicles. When asked "how safe do you think autonomous vehicle technology is compared to human drivers?", an average of 3.79 points was obtained on a 7-point scale (e.g., 1 point for high anxiety and 7 points for high safety). The result was slightly lower than 4 points, which corresponded to the same safety, indicating that the respondents thought autonomous vehicle technology was less safe than human drivers.

## 3. First field test

This section shows the major factors affecting the anxieties identified through the 1st field test. Section 3.1 shows the results of analyzing the clicker usage, and Section 3.2 shows the results of the survey on anxiety for virtual scenarios. Section 3.3 analyzes the results of in-depth interviews, and Section 3.4 summarizes the major factors that affect anxiety, based on the analysis results of previous sections. A total of 21 people participated in the 1st field test, including 3 in the 1st pilot test and 18 in the 1st main test.

### *3.1. Analysis of anxiety factors based on clicker usage results*

In the field tests, the participants used the clicker whenever they felt anxious, and the usage logs were utilized in the video interviews. During the interview, participants were told to explain why they had felt anxious, and they evaluated their degree of anxiety on a 7-point scale. Table 2 shows the anxiety factors reported by the participants in the 1$^{st}$ test, the degree of anxiety (A), the number of clicks on the clicker (B), and the number of users who pressed the clicker (C). The total score was obtained by multiplying the degree of anxiety (A) by the number of clicks (B). Based on this, the importance rankings were determined. Figure 9 gives examples of the top five factors.

[Table 2 near here]

[Figure 9 near here]

The top anxiety factors were found to be "cut-in, turning, pedestrian, illegal parking, alley, accident occurrence alarm, reckless driving (external vehicle), horn sound (external vehicle), speed, and protruding vehicle". The participants' video interview results for the major anxiety factors are as follows.

The top-ranking "cut-in" factor occurred when vehicles, taxis, and buses were cutting in around the robo-taxi. This situation most frequency occurred in congested downtown areas. Some of the participants complained about the fact that the robo-taxi allowed all cut-in vehicles, performing excessively safe driving. Below are the interview replies regarding the cut-in factor.

"Another vehicle was cutting in. I felt anxious because it was the first situation like this I had experienced in the robo-taxi." (p16)

"I felt anxious because buses and taxes cut into my lane. I might have also felt anxious if it had been a conventional taxi." (p17)

"I felt anxious because the robo-taxi allowed all cut-in vehicles. Allowing all cut-in vehicles increases the traveling time, and I thought the boarding purpose (of fast travel) could not be met." (p19)

The second-ranking "turning" factor occurred when the robo-taxi was in close proximity to nearby vehicles whilst making a U- or left turn with them. Below are the interview replies regarding the turning factor.

"I felt anxious because the robo-taxi was likely to collide with another taxi while making a U-turn." (p21)

"While the robo-taxi was making a left turn, tailgating occurred due to many vehicles on the road. If it had been a conventional taxi, a safe lane change could have been made while making a left turn." (p19)

"The robo-taxi made a left turn with vehicles nearby. I felt anxious because the safety distance was not secured." (p18)

The third-ranking "pedestrian" factor occurred when pedestrians jaywalked in congested areas, or when they stepped out suddenly at right-turn sections. Below are the interview replies regarding the pedestrian factor.

"Pedestrians jaywalked while vehicles were stuck in traffic. I was worried about whether the robo-taxi could detect the pedestrians well." (p21)

"A pedestrian jumped out in a right-turn section. I was worried about an accident." (p16)

As for the fourth-ranking "illegal parking" factor, complex situations, such as crossing the centerline to avoid illegally parked vehicles on the road, caused the participants to feel anxious. Below are the interview replies regarding the illegal parking factor.

"The robo-taxi was bypassing a vehicle parked on the side of a two-lane road by crossing the centerline. I felt worried and anxious about whether the robo-taxi could detect the nearby vehicles well or if it could cross the centerline well." (p17)

The fifth-ranking "alley" factor occurred when the robo-taxi traveled in alleyways, due to the number of vehicles parked on the sides of roads and the large numbers of pedestrians. Various obstacles present in alleyways was also a source of anxiety for the passengers. Below are the interview replies for the alley factor.

"There were many parked vehicles and pedestrians in an alley, and I was worried about an accident because the road was too narrow." (p08)

"A pedestrian was walking unaware of the robo-taxi. I was worried about a collision with the pedestrian. A function to warn the pedestrian with horns or alerts was necessary." (p13)

As described above, the top five anxiety factors were mostly caused by external situations. In particular, the passengers felt anxious about whether the robo-taxi could recognize and respond well to external vehicles and people. The sixth-ranking factor was the "accident occurrence alarm" factor. The alarm was an HMI function informing the passenger that an accident had occurred in the vicinity of a certain section. With the comment "an accident has occurred up ahead!", the accident occurrence app was executed in the passenger's display and two buttons were provided (see Figure 3.). One interesting point of the test results was that the participants used the buttons in different ways, depending on their gender; 66% of the women participants used the control tower connection button and the remaining 33% attempted a connection to the control tower, but the function did not properly operate due to a control tower communication error. On the other hand, 66% of the male participants pressed the standby button and 33% did not press any button. These results showed that women have a tendency to actively identify the current situation and to try to obtain information through communication with others in abnormal situations, whereas men have a tendency to wait for the next guidance while judging the situation by themselves. Below are the quotes taken from the interviews with participants.

"After the robo-taxi successfully passed parked vehicles and a construction site, the accident occurrence alarm app was suddenly executed and made me anxious. I was wondering about cancellation or connection to the control tower, but I pressed connection to the control tower because it was my first experience of the situation." (p17)

"I felt anxious that the accident occurrence alarm app had been executed due to a problem inside the robo-taxi. I was wondering if it was necessary to stop driving for an accident that had occurred in another vehicle. It would be better if detailed information on an accident situation were provided." (p19)

"I kept waiting, not knowing how to respond in the accident occurrence area. I don't think I would have been anxious if I had known about the buttons of the accident occurrence alarm app." (p16)

The interview results indicate that detailed information on how to use the accident occurrence alarm function and on the details of the accident situation needs to be provided, and that it is more effective for the robo-taxi to automatically select a detour method in the event of an accident in the driving path.

To determine the cause of the ninth-ranking "speed" factor, the interview contents were examined and the following results were obtained.

"The robo-taxi accelerated to keep up with other vehicles upon entering a wide road, and that made me anxious." (p04)

"The robo-taxi suddenly slowed down. I was anxious because it was my first experience of the situation." (p16)

Although the robo-taxi changed its speed after recognizing the driving environment, the participants who selected the speed factor sensitively responded to the speed change. The anxiety surrounding the speed factor appears to have occurred because it was the participants' first experience of an autonomous vehicle. It was expected that the user anxiety arising from speed will be reduced if the users have more opportunities to board robo-taxis, or if they can provide a desired speed control.

In addition, the correlation between anxiety and speed was investigated by mapping the driving speed as measured with OBD2, which was installed in the vehicle, and the degree of anxiety recorded using the clicker in the driving situations. When the average speed by section, the number of clicks, and the number of people who pressed the clicker were examined on a graph (see Figure 10), anxiety was found to be high at low speeds and low at high speeds in many cases, contrary to expectations. The highest anxiety was found in the alley section, which contained many anxiety factors despite very slow driving. Therefore, it was found that anxiety is more affected by the characteristics of each section than by the speed of the robo-taxi for driving in downtown areas.

[Figure 10 near here]

*3.2. Virtual scenario anxiety factor analysis*

In the field tests, anxiety scenarios that could not be artificially implemented could not be evaluated. Therefore, a survey on the virtual anxiety scenarios that were used in the online survey in Section 2.1 was conducted in the same way as those conducted on people who had experienced the robo-taxi. For them, more realistic evaluations were expected, despite the virtual scenarios. Table 3 shows the differences between the online respondents who did not experience the robo-taxi and the field test participants who experienced the robo-taxi. Their degree of anxiety was evaluated on a 7-point scale, and the ranking change shows the change in rankings of each anxiety factor between the evaluations of the field test participants and those of the online respondents.

[Table 3 near here]

Both the field test participants and the online respondents evaluated "when a pedestrian suddenly jumps out from a crosswalk in the driving signal" as the scenario with the highest anxiety. Of the scenarios whose rankings rose; "when the navigation system suddenly stops responding (machine malfunction) (+17)" and "when the robo-taxi drives without any explanation of the direction (+7)" were seen to rise most significantly in rankings. These scenarios are related to the malfunction of the navigation system. The field test participants could judge the error of the navigation system as the problem with autonomous driving technology, because they could predict the behavior of the vehicle only by relying on the information of the navigation system. Therefore, it is expected that the malfunction of the navigation system will significantly affect anxiety in actual robo-taxi services. Moreover, "when a vehicle approaches from the opposite direction in a narrow alley (+3)" was included in the updated top ten rankings. This appears to be because the participants had a similar experience when traveling in an alley in the field tests. On the other hand, the representative scenarios whose rankings were lowered included "when the distance to the vehicle ahead is very small (-6)" and "when the view is blocked by a large vehicle ahead (-6)". These results indicate that these are scenarios that cause less anxiety after previous experience of them, even though higher anxiety is expected before experiencing them.

In general, the anxiety evaluation scores of the field test participants were lower than those of the online respondents. This indicates that people who have experience of autonomous vehicles may feel safer about autonomous driving than those who have no experience. Therefore, it is expected that the anxiety surrounding autonomous driving can be quickly reduced if many people are allowed to experience robo-taxis at the beginning of the service. Many of the participants in the field tests also quickly adapted themselves to the robo-taxi, and felt comfortable after boarding.

### 3.3. Anxiety factor analysis through in-depth interviews

The participants were interviewed on anxiety before and after boarding the robo-taxi, and in-depth interviews were conducted on the following representative areas: (1) Concerns over the robo-taxi service, (2) shortcomings compared to conventional taxis, (3) the most anxious moments in the robo-taxi service, and (4) how to respond to errors of the robo-taxi.

#### 3.3.1. Concerns over the robo-taxi service

There were 28 overlapping answers to the question "do you have any concerns over the robo-taxi service?" Among them, 53% mentioned that there were concerns due to the lack of reliability of the technology. Many other responses mentioned the inconvenience resulting from the fact that the passenger could not control the robo-taxi, and that communication with the robo-taxi was not possible. It can be seen that the anxiety surrounding malfunctions resulting from a lack of reliability of the machine was also ranked second, fourth, sixth, and eighth in the anxiety scenarios in Table 3. Below are the quotes from the interviews.

"I think I will be anxious without a stop button that can mechanically stop the robo-taxi. I am worried that there is no device to help me handle the situation." (p02)

"I will be confused because there is nothing to communicate with in case of an emergency." (p04)

"Prompt responses are possible when an accident occurs while I am in a conventional taxi, but I was worried in the robo-taxi because there was no one to ask or communicate with in the event of an accident."(p08)

*3.3.2. Shortcomings compared to conventional taxis*

Of the 25 overlapping answers to the question "what was not good about the robo-taxi compared to conventional taxis?", 40% mentioned anxiety due to inflexible driving. Other responses included a lack of reliability, no communication with a driver, a lack of information and guidance, and no control over the vehicle operation, as seen in the previous section.

Conventional taxis can reach the destination faster, as the driver violates traffic laws to some extent if the passenger desires, and the driver generally predicts signals and traffic situations on familiar roads. The robo-taxi, however, drove within the allowed speed range and sometimes even at lower speeds than the regulations, to prevent an accident. This caused many of the participants to feel uncomfortable and anxious. In addition, they felt anxious about the lack of communication and information. The $3^{rd}$, $5^{th}$, and $7^{th}$ rankings of the anxiety scenarios in Table 3, i.e. the cases of a narrow alleyways, high-speed turning, and large objects in the middle of the road also indicated anxiety resulting from the lack of communication with the robo-taxi, or the lack of driving information. It appears that anxiety can be reduced if the corresponding information is provided and the driving behavior of the robo-taxi can be predicted before the passenger feels anxious. Below are the quotes from the interviews.

"When the robo-taxi could not exhibit the flexibility that humans do, I felt slightly uncomfortable because it traveled at lower speed than the allowed speed, even though I did not expect high speed."(p16)

"The robo-taxi could not meet its purpose because I usually take a taxi to reach the destination faster."(p01)

"I felt uncomfortable because the system did not communicate with the passenger."(p05)

"I was not feeling good when disembarking, because there was no message that the destination had been reached."(p10)

*3.3.3. The most anxious moment in the robo-taxi service*

Of the 23 overlapping answers to the question "what was the most anxious moment during the test?",

35% mentioned anxiety due to external factors, such as vehicles dangerously cutting in in front of the robo-taxi, pedestrians who jaywalked, and pedestrians who walked in front of the vehicle in a narrow alley. As the robo-taxi traveled at constant speeds, it became the target of vehicles traveling fast by frequently changing lanes, and the participants felt anxious about whether the robo-taxi could respond well to vehicles suddenly cutting in. They also felt anxious about whether the robo-taxi could recognize and respond well to pedestrians when they walked in front of the vehicle in an alley, or when they jaywalked. The participants felt uncomfortable because they could only watch such pedestrians from the inside of the vehicle, and could not communicate with the outside, for example by sounding the horn or using emergency lights. This shows that a method for the passenger to communicate with the external environment must be prepared, beyond the communication channels existing between the passenger and the robo-taxi. In the anxiety scenarios in Table 3, the $1^{st}$, $9^{th}$, and $10^{th}$ rankings were also "when a pedestrian suddenly jumps out", "when a vehicle approaches from the opposite direction in a narrow alley", and "when a motorcycle is running next to the vehicle". These also indicated anxiety from external factors. Below are the quotes from the interviews.

"Immediately after the robo-taxi started to move, the vehicle ahead started backing up. I felt anxious because it was my first time boarding a robo-taxi and because I did not know how the robo-taxi would respond to the situation. I felt uncomfortable because there were unexpected pedestrians on the road." (p04)

"Another vehicle was approaching from the opposite direction in a narrow one-way alley. I was anxious because I did not know how to respond." (p14)

"I felt anxious when other vehicles came too close or when they cut in, even though the robo-taxi operated normally. The surrounding environment made me nervous, rather than the robo-taxi." (p17)

*3.3.4. How to respond to the error of the robo-taxi*

Of the 29 overlapping answers to the question "what would you do if the robo-taxi showed abnormal actions, such as suddenly changing the driving path, or error signals?", 41% mentioned that connection to the control tower was required, and that humans are required rather than machines when a problem occurs, 38% mentioned that the vehicle should be forcibly stopped. The other responses included "calling another taxi", "driving without any action", "waiting in the vehicle", and "directly solving the problem". The need for a control tower capable of solving problems and an emergency exit protocol was raised, to reduce the anxiety of users in preparation for accidents that may occur after the commercialization of robo-taxis. There were also concerns that control towers would infringe privacy rights. Below are the quotes from the interviews.

"As there is no driver who can respond to an emergency, a control tower is required to reduce anxiety and promote a sense of security. In this case, however, passengers cannot be free from the eyes of the people in the control tower. An instruction from the control tower will be helpful before getting out of the vehicle." (p19)

"I will get out if the vehicle is stopped, but I will jump out of the vehicle if it keeps driving while continuously exhibiting abnormal behavior." (p07)

"It is necessary to stop the vehicle in a fast and safe manner. If the robo-taxi cannot recognize an unexpected situation while the passengers are aware of it, a function to safely stop driving will be required." (p18)

### *3.4. Summary of the derived anxiety factors*

Based on the analysis results of Sections 3.1, 3.2, and 3.3, the major anxiety factors can be summarized, as shown in Table 4. The top ten anxiety factors were selected based on the clicker use results, and five anxiety factors that were not implemented in the field tests were selected from among the top anxiety factors obtained through the virtual scenario evaluation results. Finally, the four fundamental anxiety factors obtained through the in-depth interviews were added.

[Table 4 near here]

## 4. HMI Design

In this section, the HMI solutions to the major anxiety factors in Table 4, derived from the 1st field test results, are proposed and designed. HMI designs resolving each anxiety factor with the simplest and minimal functions were targeted. Table 5 summarizes the final determined HMI functions, which were reflected in the 2nd field test. Among the anxiety factors, strange warning sounds, navigation system malfunction, and blinker malfunction were integrated as "abnormal operation". Table 6 shows the additional functions implemented to resolve the pain points found through the 1st field test, even though they were not functions to reduce anxiety.

[Table 5 near here]

[Table 6 near here]

Figure 11 shows the vehicle interior in which each function has been implemented. Although each function was implemented with different display devices for fast prototyping, they can be integrated into a single display device for the final product. Figure 11(a) shows the actual environment that the participants experienced when seated in the robo-taxi. The first image from the left in Figure 11(b) shows the display by which the passenger can communicate with and control the vehicle. The second and third images show displays for providing information. The detailed contents of all functions will be introduced in the following subsections.

[Figure 11 near here]

### 4.1. Speed control functions

The 1st field test results showed that many of the participants felt uncomfortable and anxious due to the excessive cruise control of the robo-taxi. To address this problem, a function was created for the participants to directly control the speed at which they felt comfortable. The robo-taxi traveled in the "default" mode, and the mode could be changed to a "fast driving" mode that enabled faster and more flexible driving when the passenger was uncomfortable with slow driving, or to a "safe driving" mode when the passenger felt anxious. The three emoticons located at the top left of the first display in Figure

11(b) are the buttons for these driving modes. Table 7 gives detailed descriptions of the three driving modes.

[Table 7 near here]

*4.2. AI voice function*

In the 1st field test, many of the participants felt uncomfortable and anxious because they could not ask and communicate with the robo-taxi. An AI speaker function was installed for communication with the robo-taxi. When the user asked a question, the control tower produced a voice response through the speaker installed inside the vehicle, using text-to-speech synthesis and the WOZ methodology instead of an actual AI speaker. Table 8 shows examples of the questions asked by the participants using the AI voice function in the 2nd field test. For reference, "Taeksong" is the name of the AI speaker.

[Table 8 near here]

Moreover, in the 1st field test, the reliability of the robo-taxi in accurately recognizing external factors, such as pedestrians, illegal parking, and obstacles, was low. Because there was no way to examine such recognition, the participants were very anxious about external factors. Therefore, a function by which the robo-taxi provides vocal assurances to the user that it has recognized external situations when it detects unsafe situations before the user does, was added to relieve the anxiety of the passenger. As shown in Table 9, insecure external situations were defined based on the experiences obtained through the 1st field test, and voice guidance for each situation was created so that such guidance could be heard whenever the corresponding situations occurred. All these forms of guidance were applied in the 2nd field test.

[Table 9 near here]

*4.3. Horn, emergency stop, direction guidance, camera, and sleep functions*

The other HMI functions employed to reduce anxiety are as follows. The first function is the horn. The participants in the 1st field test could not warn other vehicles of their incorrect driving, or pedestrians walking in front in an alley, because they could not sound the horn. Therefore, they felt anxious, uncomfortable, and embarrassed while they could not take any action. The newly provided display in Figure 11(b) allowed participants to send a signal by pressing the horn button.

The second is the emergency stop function. The participants could stop driving by pressing the stop button in the event of an emergency, or when they felt that the robo-taxi was performing abnormal driving. The passengers could take care of personal matters, such as getting off and going to a rest room. They could also stop the vehicle and get off when they wished to escape from an uncomfortable situation. When the vehicle is stopped, it departs again if the stop button is pressed again.

The third is the direction guidance function. In the 1st field test, the participants could hear the blinker sound, but they could not identify the direction. Therefore, animation was provided to the passengers to inform them of the turning direction, as shown in the middle display of Figure 11(b). This function can also be used to provide information to the passengers in the rear seats.

The fourth is the 360° camera image. The real-time 360° camera image can be provided to the passengers using the camera installed on top of the vehicle. In the interviews with the participants of the 1st field test, some feedback indicated that the participants felt anxious because they could not properly see ambulances, despite the sound coming from behind the vehicle, and because they had to look around for the ambulances. They also felt anxious when pedestrians were close to the vehicle because they were not sure about whether the robo-taxi properly recognized them. Therefore, the 360° view around the vehicle was provided to the passengers through a monitor, to reduce their anxiety.

The final function is the sleep mode. In the 1st field test, some of the participants had to endure drowsiness and respond to the test even in situations where they were not significantly anxious. The 1st field test interview results showed that "sleep" was the most desirable behavior in the robo-taxi. Moreover, in the 1st pilot test, passengers were seated in the rear seats for the test, and it was found that they felt drowsy soon and this reduced their opportunity to feel anxious. This indicated that creating an environment that allows passengers to sleep in the robo-taxi when they are drowsy, and designing a system that wakes them up at the destination, to prevent them from feeling anxious, would reduce anxiety in the overall robo-taxi experience. When the sleep mode button in Figure 11(b) was pressed, an alarm was sounded 100 m before arrival at the destination, so that passengers could sleep when they were drowsy without concern.

*4.4. Other additional functions*

Finally, the two functions in Table 6 that were added to reduce the discomfort of passengers are shown in Figure 12. The first function is the departure button. In the 1$^{st}$ field test, participants had to wait without receiving any information until the vehicle departed. It was found that waiting in the robo-taxi until something happens, in a participant's first experience of it, could promote anxiety. Therefore, the departure button was created so that the vehicle could depart when the participant desired.

The second is the robo-taxi search function. In the 1$^{st}$ field test, some participants had difficulty finding the vehicle in the boarding process. For the participants to more easily recognize the vehicle, the vehicle sounded the horn twice to indicate its location when the participant entered a 3 m radius from it and unfolded its rear-view mirrors when they entered a 1 m radius.

[Figure 12 near here]

## 5. Second field test

In this section, the anxiety reduction effects of the HMI functions designed in Section 4 are verified through the 2$^{nd}$ field test, with a robo-taxi equipped with these functions[1]. The 2$^{nd}$ field test was conducted on the same route as the 1$^{st}$ field test. Seven people participated in the 2$^{nd}$ field test (1 person in the 2$^{nd}$ pilot test and 6 people in the 2$^{nd}$ main test).

The clicker usage analysis, used in the 1$^{st}$ field test (see Table 2), was here also conducted. Table 10 shows the clicker use results of the 2$^{nd}$ field test. The change in the degree of anxiety represents a difference from the degree of anxiety in the 1st field test. For the factors common to the 1st and 2nd field tests, anxiety decreased in all case except the "alley" factor.

---

[1] To promote understanding of this study, videos of the 2$^{nd}$ pilot test and HMI function operation have been posted at the following URL: https://youtu.be/etemMpwljeg

[Table 10 near here]

In the 2nd field test, the 'inter-vehicle distance' and 'obstacle on the road' factors became apparent. The inter-vehicle distance refers to the distance between the robo-taxi carrying the participant and an external vehicle, and obstacles on the road means any animals or objects found on the road that cause anxiety. These caused anxieties surrounding collisions with external objects. The participants indicated that the drivers of conventional taxis could handle these factors well. It is judged that the robo-taxi may also reduce anxiety if it detects a collision with external factors well, or if it provides vocal assurance that it is well aware of such factors. Below are the relevant interview responses.

"When another vehicle approached, I was worried about a collision. If it had been a conventional taxi, I would have told the driver that it was too close…" (p27)

"The robo-taxi travelled fast without avoiding the many pigeons on the road. A conventional taxi would have handled the situation better. Sophisticated recognition technology or voice messages confirming recognition would be helpful." (p24)

In the 2nd field test results, 15 anxiety factors that occurred in the 1st field test did not occur. Although a simple comparison is difficult because the test was conducted with fewer participants than the 1st field test, it appears that the effects of the HMI functions were observed in light of the fact that the degrees of anxiety decreased. For more quantitative evaluation, a survey on the degrees of helpfulness of the major HMI functions in relieving anxiety was conducted on a 7-point scale, and the results are shown in Figure 13(a). All the HMI functions exhibited scores of four points or higher, corresponding to "normal", and indicating that they were helpful in reducing anxiety. In particular, direction guidance, emergency stopping, speed control, and AI voice guidance exhibited scores of five points or higher. There were some differences in the results of the survey asking what functions were required for the robo-taxi. As can be seen from Figure 13(b), the horn function was evaluated as the second necessary function for the robo-taxi. This means that the function is essential for communication with the external environment, even though it may be relatively less helpful in relieving anxiety. Moreover, after each field test, a survey on the overall satisfaction with the robo-taxi service, and the participant's willingness to use it, was conducted on a 7-point scale, and the results of the 2nd field test were compared with those of the 1st field test. As shown in Figure 13(c), the 2nd field test (with the HMI functions) exhibited higher overall satisfaction and participant willingness than the 1st field test (without HMI functions). This confirms that the HMI functions added in this study reduced anxiety, thereby providing positive robo-taxi usage experiences, and increasing willingness to use robo-taxis in the future. In addition, the vehicle search function was shown to be helpful, because it exhibited an average of 5.33 points on a 7-point scale.

[Figure 13 near here]

Along with the positive results, shortcomings were found. Although the 2nd field test generally exhibited better results than the 1st field test, some participants were dissatisfied with the fact that the robo-taxi was not as flexible in driving as conventional taxis. Furthermore, although a HMI function to communicate with passengers was added, it was not ideal for passengers who expected communication comparable to that experienced with a human driver. In addition, the 360° camera showed the lowest satisfaction, and its anxiety reduction effect was not significant. Below are the results of user interviews on the major HMI functions.

"I used various functions, and they were generally good. I feel uncomfortable with human taxi drivers, but the voice guidance of the robo-taxi was comfortable. It was possible to look around with the camera." (p27)

"Everything went well. Before arrival at the destination, I asked Taeksong (AI speaker) about the expected arrival time. Taeksong provided correct information, which was satisfactory." (p25)

"I felt safe because the robo-taxi told me that it was aware of the outside situation. I could not figure out what the 360° camera was showing. The horn function had some time delay between pressing the button and hearing the horn sound. There was no arrival guidance in the sleep mode. I think the sleep mode function is not necessary when I am anxious." (p23)

## 6. Discussion and Conclusion

### *6.1. Discussion*

The insights for reducing robo-taxi anxieties, obtained from the two field tests, are summarized as follows. They will be helpful as guidelines for future robo-taxi HMI development.

First, the robo-taxi required flexibility in its driving. In this study, the 3-level speed control functions were presented and their anxiety reduction effects were confirmed. However, a large portion of the feedback indicated that more driving flexibility is necessary, because the main purpose of using a taxi is to travel faster. In the event of an outside accident, the robo-taxi also should have been more flexible, by providing detailed information on the situation and bypassing the site. As safety and flexibility are conflicting items, they will be very important issues in future robo-taxi developments. The relevant participant interview segments are as follows.

"The purpose of taking a taxi instead of other mass transportations is to travel faster, because the passengers are in a hurry, but the autonomous driving mode could only perform normal behavior. This made me uncomfortable because human taxi drivers could travel faster, even illegally if I was in a rush." (p26)

"The robo-taxi was not good at judgments and simply stopped when there were illegally parked vehicles. I thought its judgment ability was poor and I felt uncomfortable." (p28)

Second, effective interaction with the outside is required. This is because several major anxiety factors, such as cut-ins, reckless driving, protruding vehicles, external horn sounds, and pedestrian factors, are related to external vehicles and pedestrians. The most frequent answer to the question "what was the most anxious moment during the test?" corresponded to the moment when there was an unexpected situation outside. In this study, the horn function was provided to allow the participants to communicate with the outside. Although this is the most basic function, its necessity was highly evaluated, as it was selected as the second most important function in the survey conducted on the functions required of the robo-taxi. Therefore, various methods for passengers or robo-taxis to interact with external vehicles and pedestrians need to be considered in addition to horns. Extracts from the related interviews with the participants are as follows.

"I was anxious about whether the robo-taxi could cope well with the signal violations of other vehicles." (p27)

"I was nervous because I felt the robo-taxi could not respond well to vehicles cutting in. The reaction speed was poor." (p28)

"I was worried about whether the robo-taxi could respond well to motorcycles or buses cutting in." (p24)

"A motorcycle passed by while the robo-taxi was passing through a crosswalk after turning left. I felt that the horn mode was necessary, to inform nearby vehicles of a risk." (p26)

Third, many of the participants had a low confidence in robo-taxi technology due to the risk of an accident. Therefore, it is necessary to test specific scenarios on the occurrence of accidents and faults, and to examine pain points before establishing robo-taxi services. To increase the reliability of robo-taxis and decrease anxiety, large quantities of accurate information on how to respond to unexpected situations while driving need to be provided to users. For example, it is necessary to produce guidance videos for passengers to watch before boarding robo-taxi services or while on the move, and to determine specific protocols for the responses of robo-taxis in the event of accidents. The relevant interview extracts are as follows.

"The comments, such as 'the robo-taxi had an emergency stop', 'an emergency stop will occur', and 'the robo-taxi will bypass an illegally parked vehicle', were very good. I thought urgent situations can be better handled with guidance." (p26)

"Guidance is necessary for all functions and precautions. If guidance is not possible, information needs

to be provided through in-vehicle voice guidance." (p25)

"Anxiety will be reduced if descriptions on several specific situations and how to respond to them are provided to passengers before boarding the robo-taxi." (p24)

Fourth, basic functions required to relieve robo-taxi anxiety could be confirmed in this study. The results of this study showed that robo-taxis must provide not just the information that would be provided by human taxi drivers, but also extra information to reduce anxiety. Seven major HMI functions were designed in this study. Among them, the direction guidance, emergency stop, speed control, and AI voice guidance functions exhibited relatively large effects on anxiety relief. It was found that a horn function must be included as a basic function of robo-taxis, to facilitate communication with external environments. In particular, speed control, AI voice guidance, and horn functions have not been implemented in currently available robo-taxi services, such as Waymo. They need to be reflected in future robo-taxi services (The Verge, 2018). The navigation system is the most basic function of the robo-taxi, but its malfunction could provide relatively high anxiety because it is regarded as a problem with autonomous driving technology. The relevant interview extracts are as follows.

"I was not using the fast driving function because I was worried about an accident, but the speed was appropriate. The stop and horn functions were satisfactory, but the 360° camera images were complex and dizzying. I was also satisfied with the AI voice guidance function." (p25)

"The fast driving mode was good because it was faster and safer than I thought." (p22)

In addition to the major insights described above, several interesting phenomena could also be found through the field tests. First, differences in behavior between men and women in the event of an accident were found through the accident occurrence alarm app scenarios. While women showed a tendency to aggressively identify the current situation by communicating with real people, men tended to wait for the next guidance while they further observed and judged the situation.

Second, there was no correlation between speed and anxiety. When the relationship between the clicker usage record and the vehicle speed was examined, it was found that the anxiety of the users was caused by surrounding elements, and that the sections where they had felt high anxiety had a low correlation with the vehicle speed.

*6.2. Conclusion*

With the accelerating commercialization of robo-taxi services, there is a growing need for research on the anxieties produced in potential customers by autonomous vehicles. To relieve such anxieties, it is necessary to provide solutions through the improvement of autonomous driving technologies, and to identify the fundamental causes of the anxieties. If these factors can be resolved using uncomplicated HMI methods, the methods will contribute to the stable establishment of robo-taxi services. The purpose of this study was to provide robo-taxi usage experiences of real road situations in complex downtown areas, to identify the factors responsible for anxiety during such experiences, and to provide HMI solutions to resolve such factors. A robo-taxi service that can be safely tested in downtown Seoul was implemented using the Wizard of Oz (WOZ) methodology, and the one-hour driving distance was tested with 28 participants to evaluate various anxiety occurrence scenarios. From the 1$^{st}$ field test, 19 major anxiety factors were derived, and seven HMI functions were designed to resolve the factors. The effects of the HMI functions were verified in the 2$^{nd}$ field test.

The anxiety factor analysis results – based on the customer experiences obtained through the field tests – are expected to be used as guidelines in HMI design for robo-taxi services in the future. In addition, various ways of resolving the major anxiety factors identified will be created, and it will be possible to develop better HMI concepts for anxiety relief, by benchmarking the HMI concept presented in this study. Finally, the WOZ methodology implemented in this study can be used in various robo-taxi field tests.

The following future studies will be conducted. First, intensive user experience (UX) evaluation and solution development will be performed for specific scenarios presenting a high customer demand. For example, it was found that the demand of female consumers for the late-night use of robo-taxis was high. This is because robo-taxis carry no risks of the crimes sometimes committed by human taxi drivers. Moreover, UX solutions are required for autonomous vehicles to be adequately prepared for accidents.

In particular, as it is likely that those from socially underprivileged groups, such as the elderly, children, and people with disabilities, will board robo-taxis alone, solutions are required for them to easily cope with accidents.

Second, UX evaluation is required for specific HMI functions. For example, in the results of this study, there was a high customer need for interaction with external vehicles and pedestrians. Effective interaction solutions and specific HMI solutions to improve kindness and friendliness need to be developed, and UX evaluation for such solutions is required.

Third, an upgrade of HMI functions is required for commercialization. For example, satisfaction with the sleep mode HMI was not high in this study, because it consisted of only an alarm function. When the function is applied to real vehicles, however, it will be possible to design better sleep environments, by using automatically tilting seats, reducing the illuminance, installing automatic sun shades on windows, playing autonomous sensory meridian response (ASMR) videos, and burning sleep-inducing incense.

**Acknowledgements**

This work was supported by Hyundai Motor Company and the National Research Foundation of Korea (NRF) grants funded by the Korean government (No.2017R1C1B2005266 and No.2018R1A5A7025409). The authors wish to thank Eunju Jeong, Ah-hyeon Jin, Seohui Joung, Woojin Kwak, Gyuwon Lee, Jihyun Lee, Yunha Park, Hanyoung Ryu and Seungyeon Shin, who were undergraduate intern researchers, for their support in setting the tests, conducting surveys, and summarizing the survey results.

**Table 1. Anxiety scenario rankings and their application to service sections**

| Anxiety ranking | Section | Virtual scenario |
|---|---|---|
| 1 | S1 | When a pedestrian suddenly jumps out from a crosswalk in the driving signal |
| 2 | S1 | When a large object falls from a truck in the middle of the road |
| 3 | S3 | When turning at high speed without slowing down |
| 4 | S5 | When traveling at high speed in a narrow alley |
| 5 | S2 | When the distance to the vehicle ahead is very small |
| 6 | S4 | When turning left with the right blinker on (malfunction) |
| 7 | S4 | When there is a strange noise in the vehicle |
| 8 | S2 | When the view is blocked by a large vehicle ahead |
| 9 | S1 | When a motorcycle is running next to the vehicle |
| 10 | S4 | When stopping in the middle of a crosswalk |
| 11 | S2 | When trying to change lanes in a congested area |

**Table 2. Results of deriving anxiety factors based on the clicker**

| Rank | Anxiety factor | Total score (A × B) | Degree of anxiety (7-point scale) (A) | Number of clicks (B) | Number of users who pressed the clicker (C) |
|---|---|---|---|---|---|
| 1 | Cut-in | 58.5 | 3.25 | 18 | 11 |
| 2 | Turning (left-turn/right-turn/U-turn) | 42.5 | 4.25 | 10 | 8 |
| 3 | Pedestrian | 41.5 | 3.78 | 12 | 10 |
| 4 | Illegal parking | 40.0 | 3.33 | 12 | 11 |
| 5 | Alley/narrow road | 33.5 | 4.79 | 8 | 8 |
| 6 | Accident occurrence alarm | 27.0 | 4.50 | 6 | 6 |
| 7 | Reckless driving/overtaking vehicle | 26.5 | 4.42 | 7 | 6 |
| 8 | Horn sound (external vehicle) | 24.5 | 3.50 | 7 | 5 |
| 9 | Speed (fast or slow) | 24.5 | 4.10 | 6 | 5 |
| 10 | Protruding vehicle | 24.0 | 4.00 | 6 | 5 |
| 11 | Backing vehicle (external vehicle) | 21.0 | 4.20 | 5 | 5 |
| 12 | Congested area | 21.0 | 4.20 | 5 | 5 |
| 13 | Lane change | 19.5 | 3.90 | 6 | 4 |
| 14 | Lack of information (getting off, stop, and detour) | 15.0 | 3.00 | 5 | 4 |
| 15 | System errors (GPS error and traveling path error) | 13.5 | 3.40 | 4 | 3 |
| 16 | Vehicle delay | 10.5 | 3.50 | 3 | 3 |
| 17 | Joining section | 10.0 | 5.00 | 2 | 1 |
| 18 | Interrupted view | 7.5 | 3.75 | 2 | 2 |
| 19 | Motorcycle | 7.0 | 4.50 | 2 | 2 |
| 20 | Excessive information | 7.0 | 3.50 | 2 | 2 |
| 21 | Driving without flexibility (compliance with stop lines and traffic laws) | 7.0 | 3.50 | 3 | 3 |
| 22 | Lack of adaptation to new technology | 5.0 | 2.50 | 2 | 2 |
| 23 | Uphill road | 2.0 | 2.00 | 1 | 1 |
| 24 | Sudden stop | 1.5 | 1.50 | 1 | 1 |
| 25 | Ground imbalance | 1.0 | 1.00 | 1 | 1 |
| 26 | Driving at a yellow signal | 1.0 | 1.00 | 1 | 1 |

**Table 3. Anxiety virtual scenario comparison**

| Rank | Virtual scenario | Field test participants | Online respondents | Ranking change |
|---|---|---|---|---|
| 1 | When a pedestrian suddenly jumps out from a crosswalk in the driving signal | 6.00 | 6.59 | 0 |
| 2 | When there is a strange warning sound in the vehicle while driving (When the sound is heard whenever the driver is anxious) | 5.71 | 5.72 | +5 |
| 3 | When traveling at high speeds in a narrow alley | 5.57 | 5.99 | +1 |
| 4 | When the navigation system suddenly stops responding (navigation system malfunction) | 5.56 | 4.45 | +14 |
| 5 | When turning at high speed without slowing down | 5.40 | 6.09 | -2 |
| 6 | When turning left with the right blinker on (machine malfunction) | 5.36 | 5.76 | 0 |
| 7 | When a large object falls from a truck in the middle of the road | 5.21 | 6.22 | -5 |
| 8 | When stopping in the middle of a crosswalk | 5.14 | 5.11 | +2 |
| 9 | When a vehicle approaches from the opposite direction in a narrow alley | 5.00 | 4.93 | +3 |
| 10 | When a motorcycle is running next to the vehicle | 4.96 | 5.13 | -1 |
| 11 | When the distance to the vehicle ahead is very small | 4.86 | 5.86 | -6 |
| 12 | When trying to change lanes in a congested area | 4.82 | 5.04 | -1 |
| 13 | When the robo-taxi (driver) drives without any explanation of the direction | 4.82 | 4.32 | + 7 |
| 14 | When the view is blocked by a large vehicle ahead | 4.50 | 5.42 | - 6 |
| 15 | When crossing an intersection with blinking yellow traffic lights | 4.50 | 4.85 | - 2 |
| 16 | When traveling on a road under construction without lanes | 4.46 | 4.34 | +3 |
| 17 | When nearby vehicles travel at high speed | 4.36 | 4.82 | -3 |
| 18 | When traveling on unfamiliar roads | 4.33 | 4.50 | -1 |
| 19 | When an ambulance is following with a siren | 4.32 | 4.62 | -4 |
| 20 | When going down a steep area | 4.25 | 4.60 | -4 |
| 21 | When turning left or right at a crosswalk | 3.93 | 4.20 | +1 |
| 22 | When trying to make a U-turn | 3.68 | 4.28 | -1 |

**Table 4. Derived anxiety factors (19 factors)**

| Analysis method | Major anxiety factors | Remark |
|---|---|---|
| Clicker use results (Section 3.1) | Cut-in, turning, pedestrian, illegal parking, alley, accident occurrence alarm, reckless driving (external vehicle), horn sound (external vehicle), speed, and protruding vehicle | Top ten factors |
| Virtual scenario evaluation (Section 3.2) | Strange warning sound, navigation system malfunction, obstacle on the road, vehicle stop in the middle of a crosswalk, and blinker malfunction | Top five factors that did not occur in the field tests |
| In-depth interviews (Section 3.3) | Lack of technical reliability, communication with external environments, lack of information, and robo-taxi out of control | Four fundamental anxiety factors |

**Table 5. Major HMI solutions**

| No | HMI | Detailed function | Anxiety factors that can be resolved |
|---|---|---|---|
| 1 | Speed control | The passenger can adjust speed in three steps | Fast or slow speed / turning / alley / robo-taxi out of control |
| 2 | AI voice guidance | Guidance after detecting anxious situations / Response to passenger questions / Providing diverse information | Cut-in / pedestrian / illegal obstacle / alley / abnormal operation / reckless driving / protruding vehicle / stopping in the middle of a crosswalk / lack of technical reliability |
| 3 | Horn function | Warning or alerting pedestrians / Warning or alerting nearby vehicles | Protruding vehicle / horn sound / reckless driving / alley / illegal parking / obstacle on the road / pedestrian / communication with external environments |
| 4 | Emergency stop | The passenger can stop the vehicle forcibly / Driving can be re-started when desired by the passenger | Abnormal operation / robo-taxi out of control |
| 5 | Direction guidance | Notifying turning directions in advance | Turning / lack of information |
| 6 | 360° camera view | Providing a real-time 360° view around the vehicle | Alley / pedestrian / lack of information |
| 7 | Sleep mode | Waking the passenger up at the desired time when the destination is reached | Overall anxiety relief |

**Table 6. Other additional functions**

| No | HMI | Detailed function |
|---|---|---|
| 1 | Departure button | After boarding, the vehicle can depart when the passenger desires. |
| 2 | Vehicle search | When the passenger approaches the robo-taxi they called, it will sound the horn or unfold mirrors. |

**Table 7. Details of the speed control functions**

| Function name | Driving characteristics | Application situations |
|---|---|---|
| Fast driving | • Maximum speed: 60 km/h.<br>• Inter-vehicle distance when stopped: 1 m or less<br>• Inter-vehicle distance while driving: 1-3 m<br>• For departure from the stop, speed is rapidly increased. After quickly detecting a congested area, pass the area through lane changes. | When uncomfortable with slow robo-taxi driving (ex. on roads without vehicles and for faster arrival at the destination). |
| Default | • Maximum speed: 50 km/h. Inter-vehicle distance when stopped: 1 m.<br>• Inter-vehicle distance while driving: 3-5 m<br>• For departure from the stop, speed is slowly and smoothly increased.<br>• Perform safe driving after recognizing all situations. | When stable speed is desired<br>When not in a rush<br>(ex. general road conditions) |
| Safe driving | • Maximum speed: 30 km/h. Inter-vehicle distance when stopped: 2 m or longer<br>• Safe driving: The inter-vehicle distance of 4-7 m to be maintained while driving<br>• Almost no lane change while driving. For departure from the stop, speed is increased very slowly.<br>• Yielding and defensive driving is performed as much as possible. Very slow driving in alleys. | When very nervous and anxious<br>(ex. in alleys and on market streets) |

**Table 8. Examples of AI voice communication**

| Function name | Examples of voice questions in the 2nd field test | Application situations |
| --- | --- | --- |
| **AI speaker (aka Taeksong)** | "Taeksong, how is the weather today?"<br>"Taeksong, turn on the radio"<br>"Taeksong, turn on the heater"<br>"Taeksong, activate the sleep mode"<br>"Taeksong, when will we arrive?"<br>"Taeksong, why are we stuck in traffic?"<br>"Taeksong, is there any problem with the car?"<br>"Taeksong, why did you stop here?" | - When travel-related information is required<br>- When it is necessary to control the basic functions of the vehicle (e.g., opening windows and turning on the heater)<br>- When questions are asked out of boredom |

**Table 9. AI voice guidance by situation**

| Situation | Voice guidance |
|---|---|
| Pedestrian: When pedestrians walk or jump out in front of the vehicle | "The robo-taxi has recognized pedestrians. Rest assured." |
| Narrow path: When passing through alleys or very narrow roads due to parked vehicles | "The robo-taxi has recognized a narrow alley. It will drive safely." |
| Traffic congestion: When violating a centerline in a congested area due to large numbers of vehicles, after turning left/right | "A congested area has been recognized. The robo-taxi will drive carefully." |
| Illegal parking: When violating a centerline or a lane while attempting to pass vehicles illegally parked on the sides of a road | "Illegal parking has been recognized. The robo-taxi will drive carefully." |
| Sudden stop: When a sudden stop is required due to the traffic signal or external vehicles (reckless driving) | "The robo-taxi made a sudden stop. Are you all right?" |
| Sleep mode: When the sleep mode is in operation. | "300 m to the destination. Please wake up, bang bang!" |
| Recommendation on the use of the function: When the user is bored and does not use the provided function. | "Various modes are available. Please give them a try as the robo-taxi will drive safely." |
| On departure | "Hello, welcome aboard the robo-taxi." "Please fasten your seat belt for safety." "The robo-taxi's destination is the entrance to Gyeonggyojang." "Please press the departure button when ready." |
| On arrival | "100 m to the destination. Please check if all your belongings are with you." "When the door is open, please look outside before leaving the vehicle.", "Thank you for using the robo-taxi. See you again next time." |

**Table 10. Comparison of the 1st and 2nd field test results through the clicker**

| Rank | Anxiety factor | Total Score (A×B) | Degree of anxiety (7-point scale)(A) | Number of clicks (B) | Number of people who pressed the clicker (C) | Change in the degree of anxiety |
|---|---|---|---|---|---|---|
| 1 | Cut-in (external vehicle) | 16.0 | 3.20 | 5 | 2 | -0.05 |
| 2 | Inter-vehicle distance | 14.5 | 4.83 | 3 | 2 | New |
| 3 | Alley | 15.5 | 5.20 | 3 | 3 | +0.40 |
| 4 | Sudden stop | 7.5 | 3.75 | 2 | 2 | -2.25 |
| 5 | Left turn | 4.0 | 4.00 | 1 | 1 | -0.25 |
| 6 | Inflexible driving | 3.0 | 3.00 | 1 | 1 | -0.50 |
| 7 | Pedestrian | 2.0 | 2.00 | 1 | 1 | -1.70 |
| 8 | Obstacle on the road | 5.0 | 5.00 | 1 | 1 | New |
| 9 | Slow driving | 3.0 | 3.00 | 1 | 1 | -1.80 |

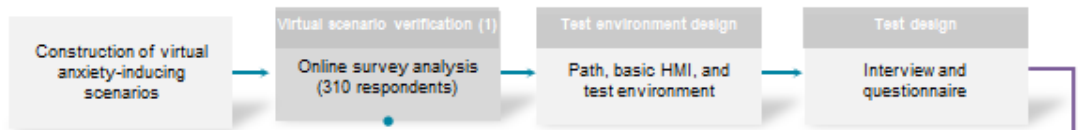
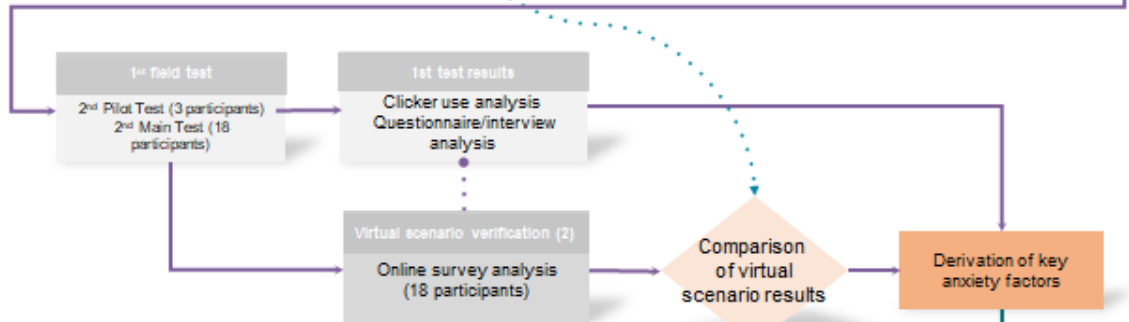
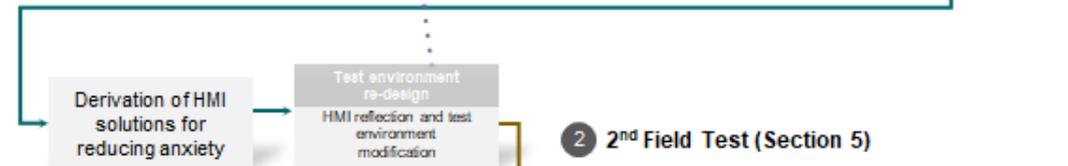
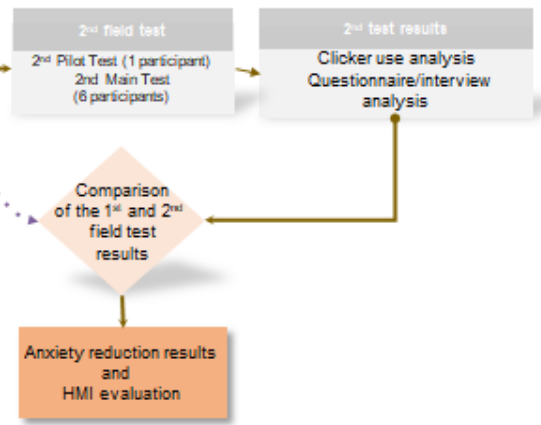

**Figure 1. Research Framework**

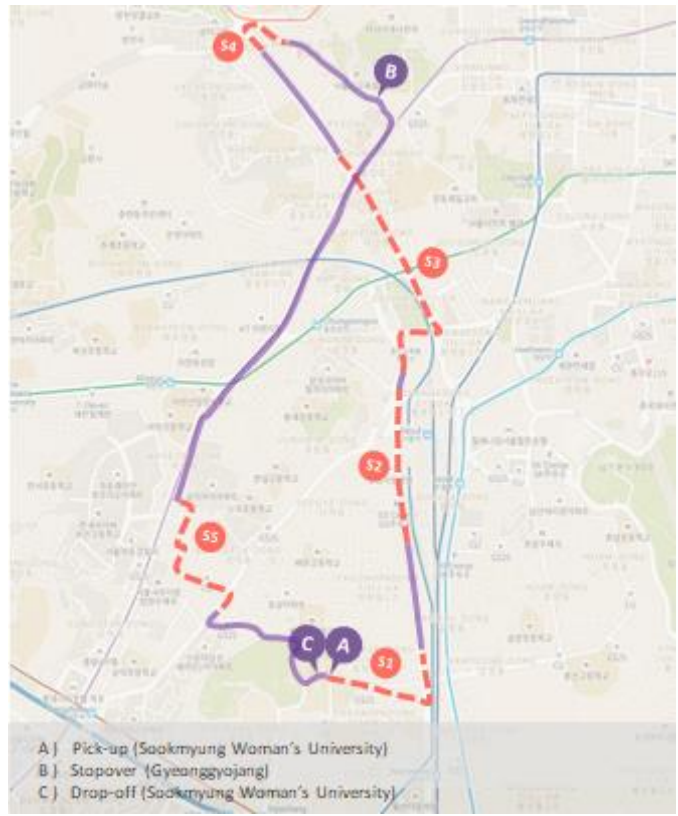

**Figure 2. Robo-taxi service path (Yongsan-gu, Seoul)**

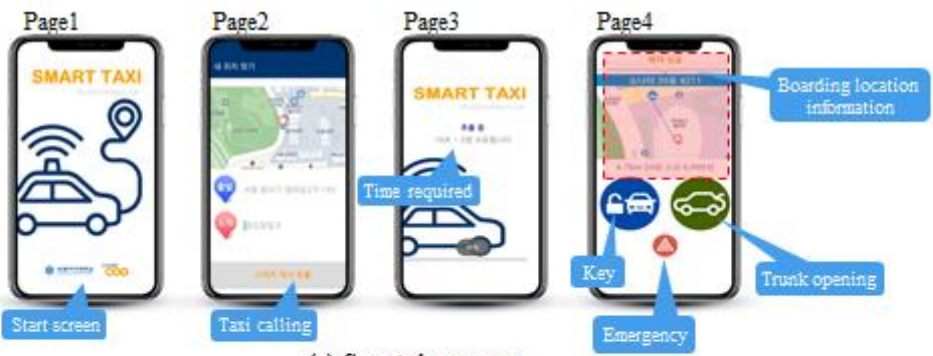

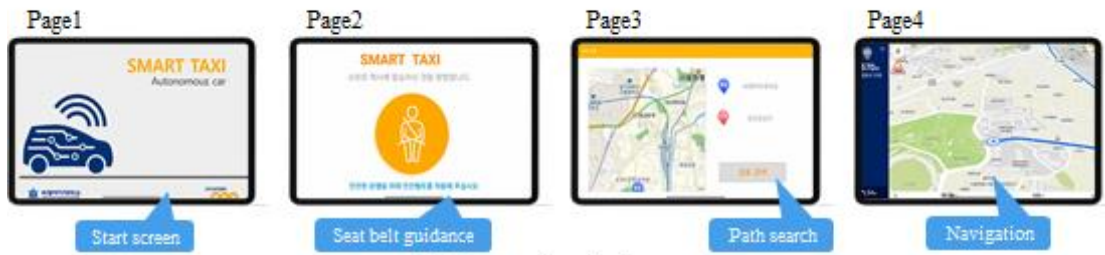

**Figure 3. Smartphone Application and Interaction Display**

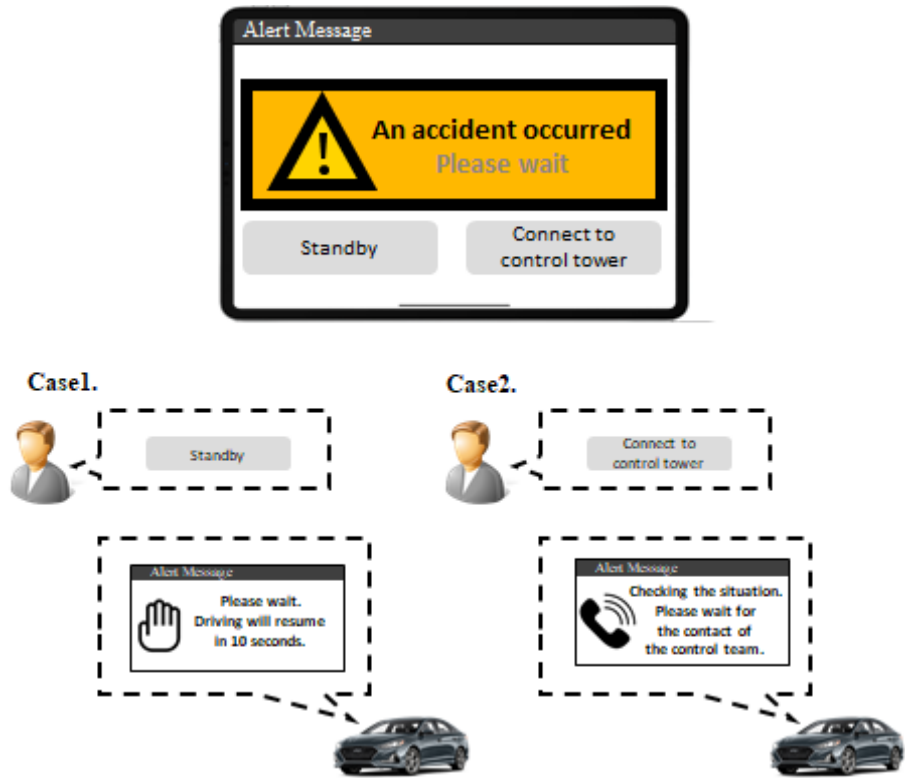

**Figure 4. Accident occurrence alarm function**

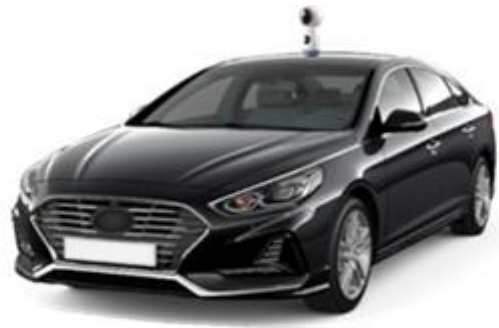

a. Exterior

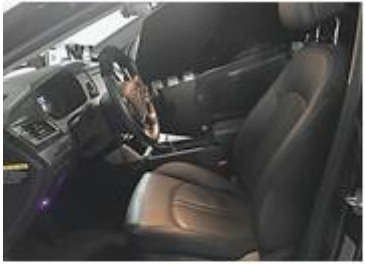 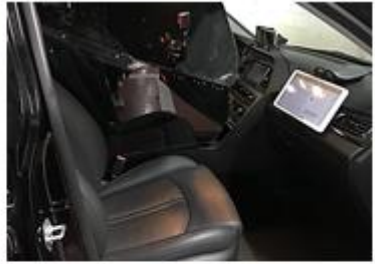 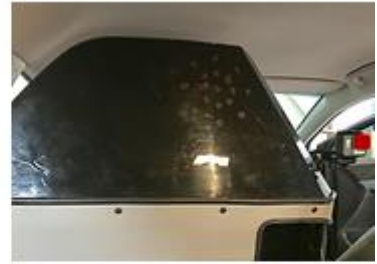

(a) Driver's seat    (b) Passenger seat    (c) View from behind the driver's seat

b. Interior (1st test)

**Figure 5. Robo taxi exterior/interior environment**

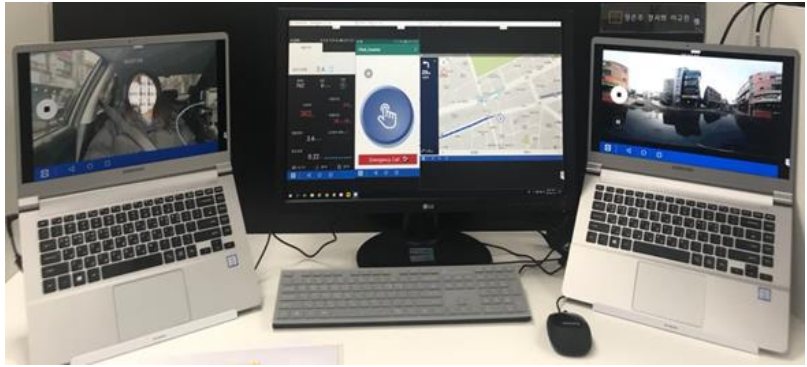

**Figure 6. Control tower environment in the 1st field test**

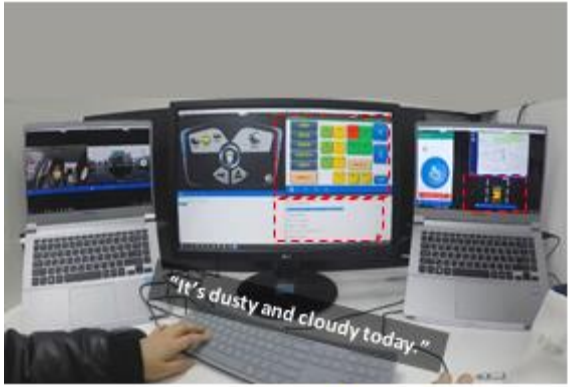 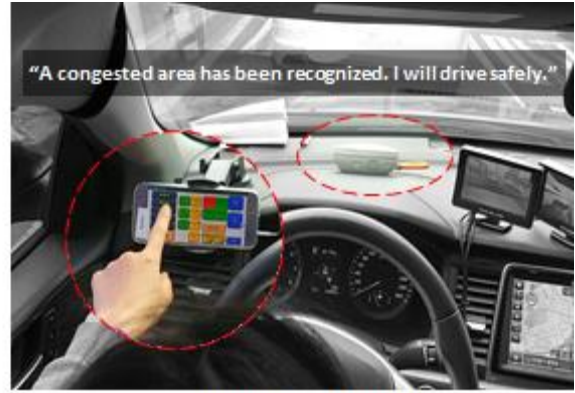

**Figure 7. Control tower and driver's seat environments in the 2nd field test**

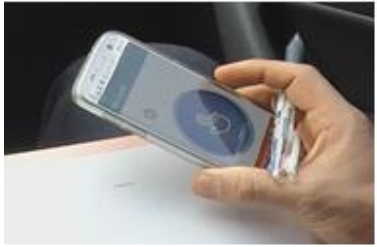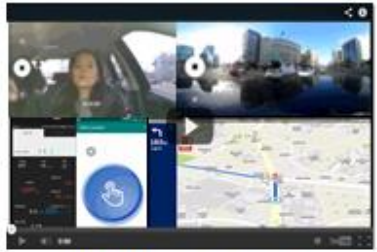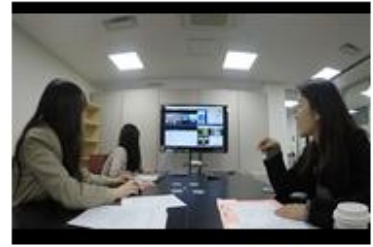

**Figure 8. Example of clicker and video interview**

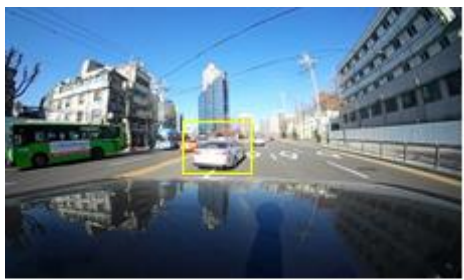 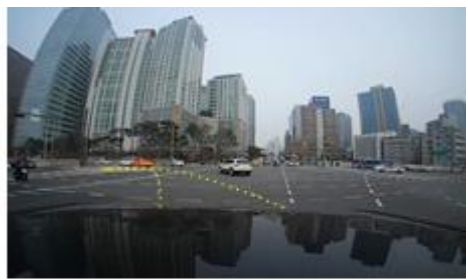 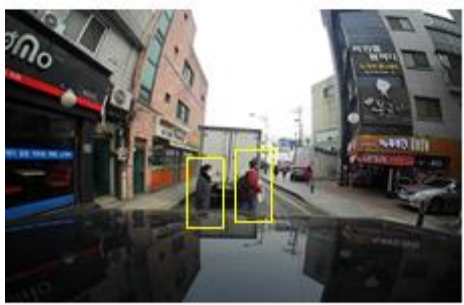 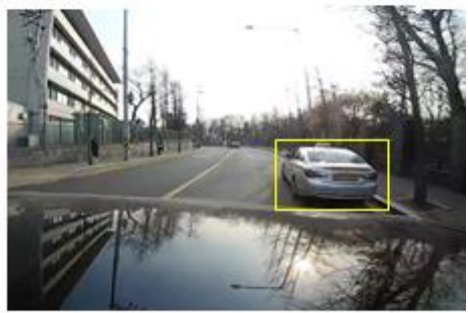 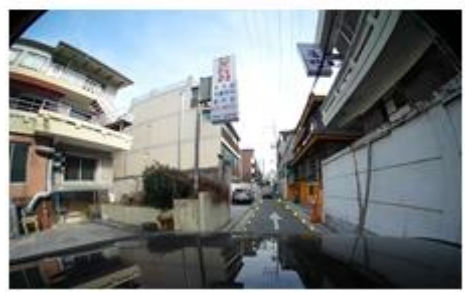

**Figure 9. Examples of the major anxiety factors**

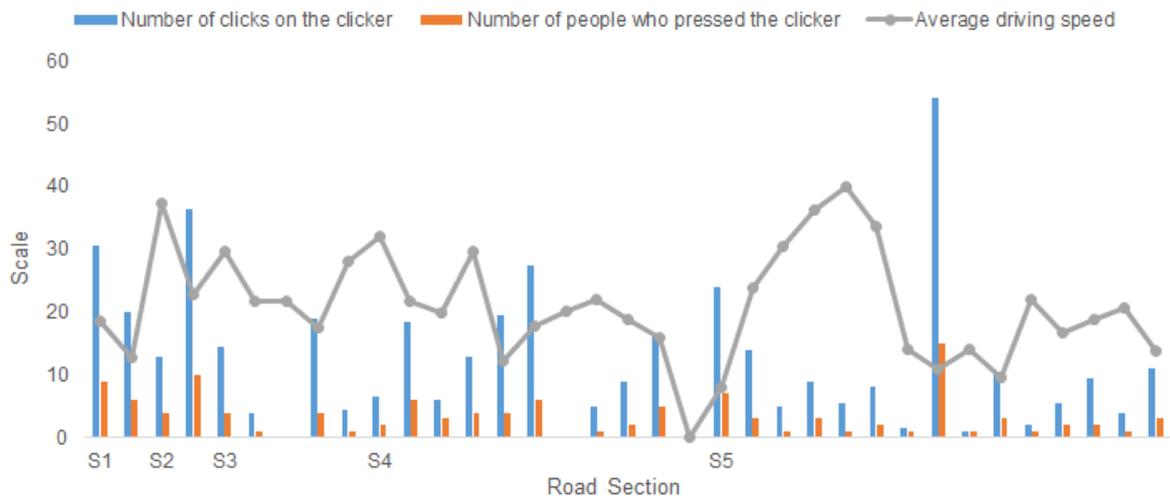

**Figure 10. Correlation between speed and anxiety**

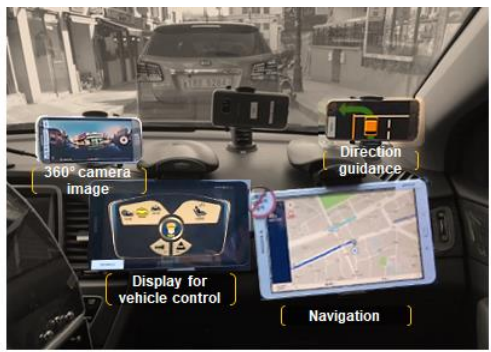

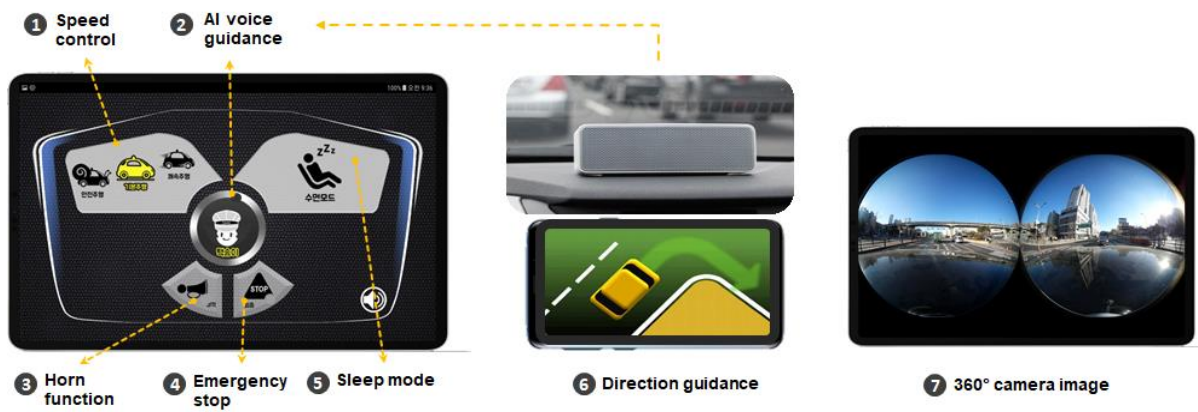

**Figure 11. HMI Solution**

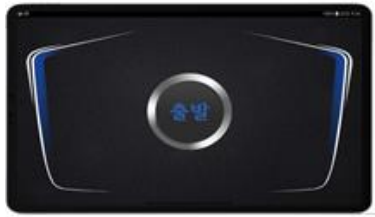 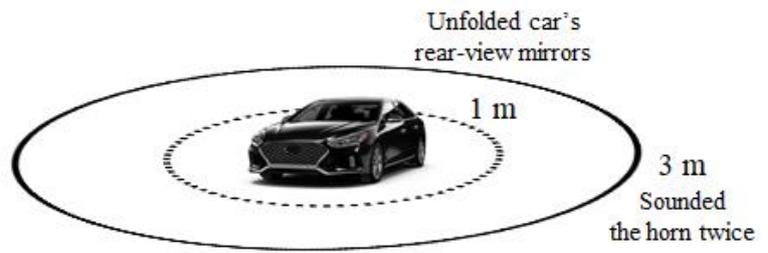

Figure 12. Other HMI functions

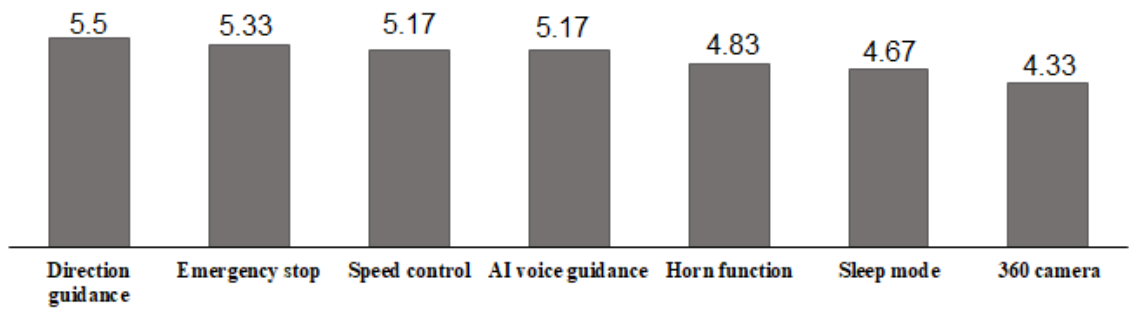

(a) Degrees of helpfulness of HMI functions in relieving anxiety

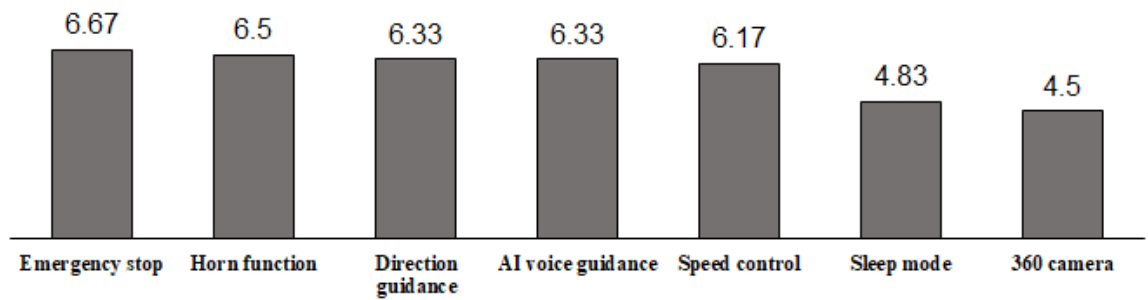

(b) Functions required for the robo-taxi

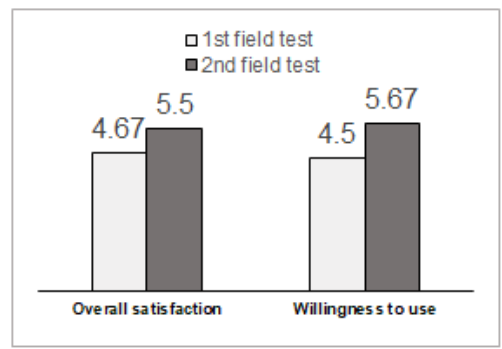

(c) Comparison of satisfaction and willingness to use

**Figure 13. HMI effects**